\documentclass[aps,apl,preprint]{revtex4}
\usepackage{amsmath,amsfonts,amssymb,graphicx,epsfig}
\usepackage{bbold}
\usepackage{hyperref}

\begin{document}

\title{\bf Topological cutoff frequency in a slab waveguide: Penetration length in topological insulator walls}
\author{T. M. Melo, D. R. Viana, W. A. Moura-Melo, J. M. Fonseca and A. R. Pereira \\  
  $ ^1$ Departamento de F\'{\i}sica, Universidade Federal de Vi\c{c}osa \\ 36570-000, Vi\c{c}osa, Minas Gerais, Brazil.\\
}

\begin{abstract}
We study the propagation of electromagnetic (EM) waves in a slab-type waveguide which walls consist of three-dimensional topological insulator (3D TI). The results show that a cutoff frequency with topological stability limits the spectrum that propagates along the waveguide and are in agreement with experimental observations. Our approach also provides a way to measure the penetration length of surface metallic states in 3D TI.  
\end{abstract}


\noindent Corresponding author: Thiago M. Melo\\
E-mail: thiago.melo@ufv.br

\maketitle

\newpage

\section{Introduction}
\label{}

The spin quantum Hall (SQH) electronic state was discovered experimentally in the HgTe/CdTe quantum wells being called two-dimensional topological insulator (2D TI) \cite{moore,zhang,koniq}. Subsequently, this electronic phase was found in materials such as $Bi_2 \, Te_3 $ and $ Bi_2\, Se_3 $, named 3D TI \cite{hsieh,xia}. They are characterized by a full insulating gap in the bulk and gapless metallic surface (edge for 2D TI) states which are protected by time reversal symmetry (TRS) \cite{CZhang}. These surface states remain robust against small non-magnetic perturbations and they are characterized by topological invariants that are preserved as long as hamiltonian varies smoothly \cite{L Fu}. In addition to their fundamental interest, their special properties make these states useful for several applications, from spintronics to quantum computation and optics \cite{lfu,aki,ychen,xzhang}.

The long wavelength regime of EM wave of 3D TI may be described by a topological field theory (TFT) whose coefficients are quantized in terms of the fine structure constant, $\alpha ={e^2/\hbar c}$ \cite{A Karck}. This TFT describes the topological magnetoelectric effect (TME) on the surface of the material when the TRS is broken at the surface \cite{XLQi and SCZhang}. Thus, in the presence of EM fields, an electric field induces a magnetization whereas a magnetic field induce an electric polarization. This topological effect is caused by a quantized Hall current induced on the surface of the TI in response to an electric field, in (3+1) D, the TFT is given by the action below \cite{indmon}:
\begin{equation} \label{S}
\mathcal{S}=\frac{1}{8\pi}\int (\epsilon \vec{E}^2-\frac{1}{\mu} \vec{B}^2) \,d^{4}x+\frac{\theta}{2\pi}\frac{\alpha}{2\pi}\int \vec{E}\cdot\vec{B}\,d^{4}x,
\end{equation}
where $\vec{E}$ and $\vec{B}$  are the electric and magnetic fields respectively, $\mu$ and $\epsilon$ are the permeability and permissivity of the TI and $\theta$ is the relevant topological parameter accounting for the TME effect and assumes $\theta = 0$ for an ordinary insulator and $\theta =\pi$ for TI. This effective theory is valid only if a surface gap is open on the TI, this can be done by covering its surface with a thin magnetic film or applying an external magnetic field perpendicular \cite{qizhang}. Among other consequences, when light is send to a TI border its behaviour is deeply modified, for example, the light polarization plane is rotated by a universal angle leading to changes in the usual Kerr and Faraday effects \cite{Top Quant}.

\section{Slab waveguide with topological insulator walls}

\begin{figure}[h]
\begin{center}
\resizebox{!}{5.5cm}{\includegraphics{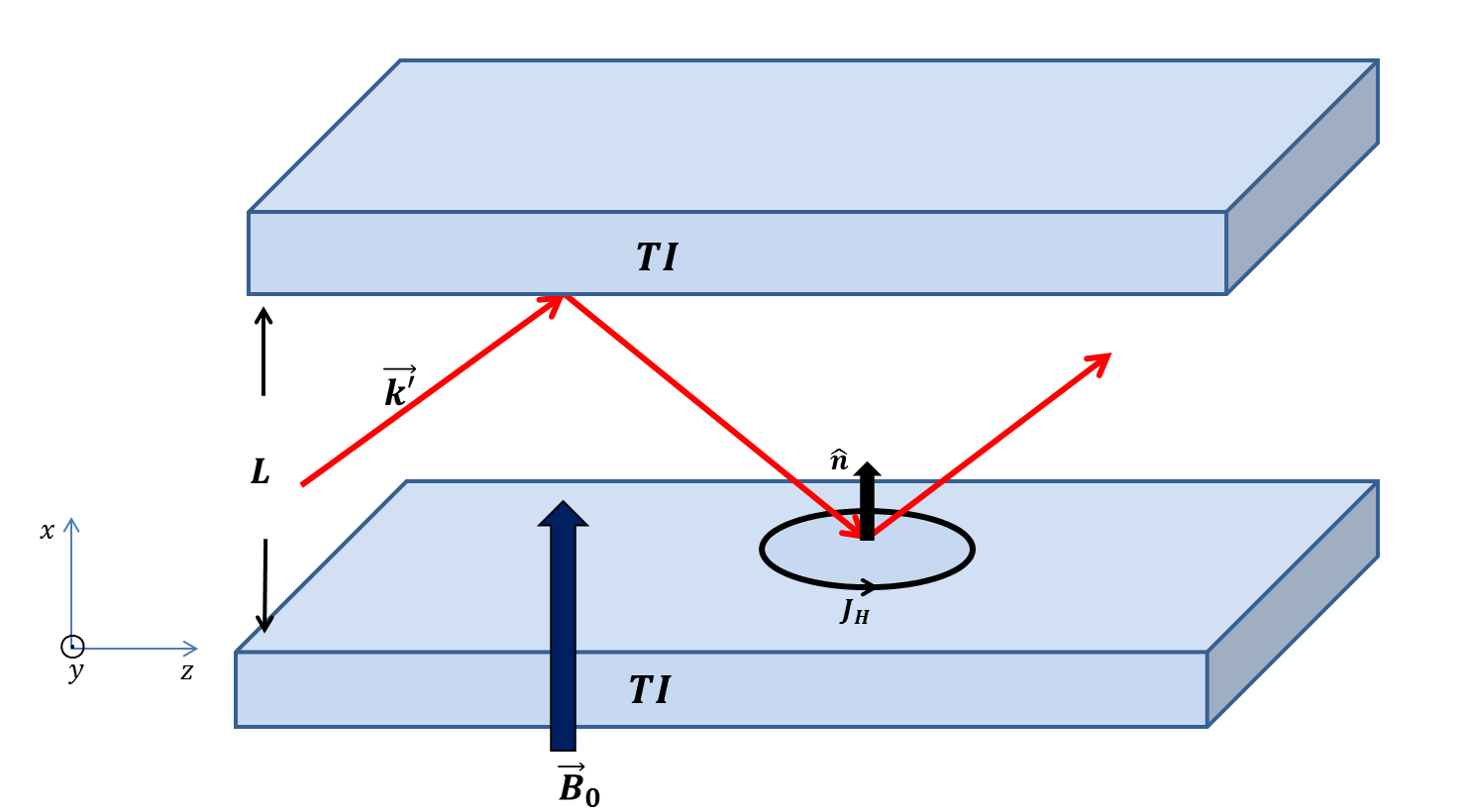}}
\end{center}
\caption{\small Slab waveguide with aperture length $L$ and TI walls. In $0<x<L$ the slab is filled with an ordinary dielectric. $\vec{k'}$ is the wave vector of a mode allowed by the system, $J_{H}$ is the Hall current induced by EM waves and circulates along the interface vacuum/TI. $\hat{n}$ is an unitary vector normal to TI surface and gives the direction of the   external magnetic field applied ($\vec{B}_{0}$).} 
\label{fig1}
\end{figure}

Here, we study the propagation of EM radiation confined to a slab-type waveguide which walls are the surface borders of a 3D TI. The interior of the guide is filled with an usual dielectric (vacuum, for simplicity), with this, there is a variation in $\theta$ at the interface of the TI and thus a contribution from topological term in Eq. (\ref{S}). We consider an aperture $L$ along $x$ and an infinite extension along $y$ and $z$ directions. In addiction, an external magnetic field is applied perpendicularly to the walls, as schematically drawn in Fig.\ref{fig1} . 

The functional variation of Eq. (\ref{S}) yields Maxwell's equations in the presence of TME, that read:
 \begin{equation}  \label{Gauss}                                                              
\vec{\nabla}\cdot \vec{E}=\frac{\rho_{H}}{\epsilon}\,                                                                    
                              \end{equation}   
\begin{equation}\label{no mon}                                        
\vec{\nabla} \cdot \vec{B}=0,                                  
\end{equation}  
    
\begin{equation}\label{Faraday}                                
\vec{\nabla} \times \vec{E}=-{1\over c} \partial_t \vec{B},  
\end{equation}
                               
\begin{equation}\label{Ampere-Maxwell}     
\vec{\nabla} \times \vec{B}= \mu \vec{J}_{H}+ {\mu  \epsilon \over c} \partial_t \vec{E}.     
\end{equation}
Most studies on TME consider the charge ($\rho_{H}$) and Hall current ($\vec{J}_{H}$) densities perfectly concentrated on TI surface using a Dirac $\delta(\vec{r})$ distribution. We take into account the penetration length, $l$, of surface metallic states on TI bulk. Thus, in our approach: $\rho_{H}=-\frac{\alpha  \theta}{\pi l} (\hat{n} \cdot \vec{B})$ and $\vec{J}_{H}=\frac{\alpha  \theta}{\pi l} (\hat{n}  \times \vec{E})$, where $l \sim 10^{-9} m$ (few nanometers) for usual 3D TI, like $ Bi_2\, Se_3 $ \cite{WZhang RYu,jacob}. While $\hat{n}$ is a unitary vector perpendicular to the walls of the waveguide.

\section{Results and discussion}

\begin{figure}[h]
\begin{center}
\resizebox{!}{5.9cm}{\includegraphics{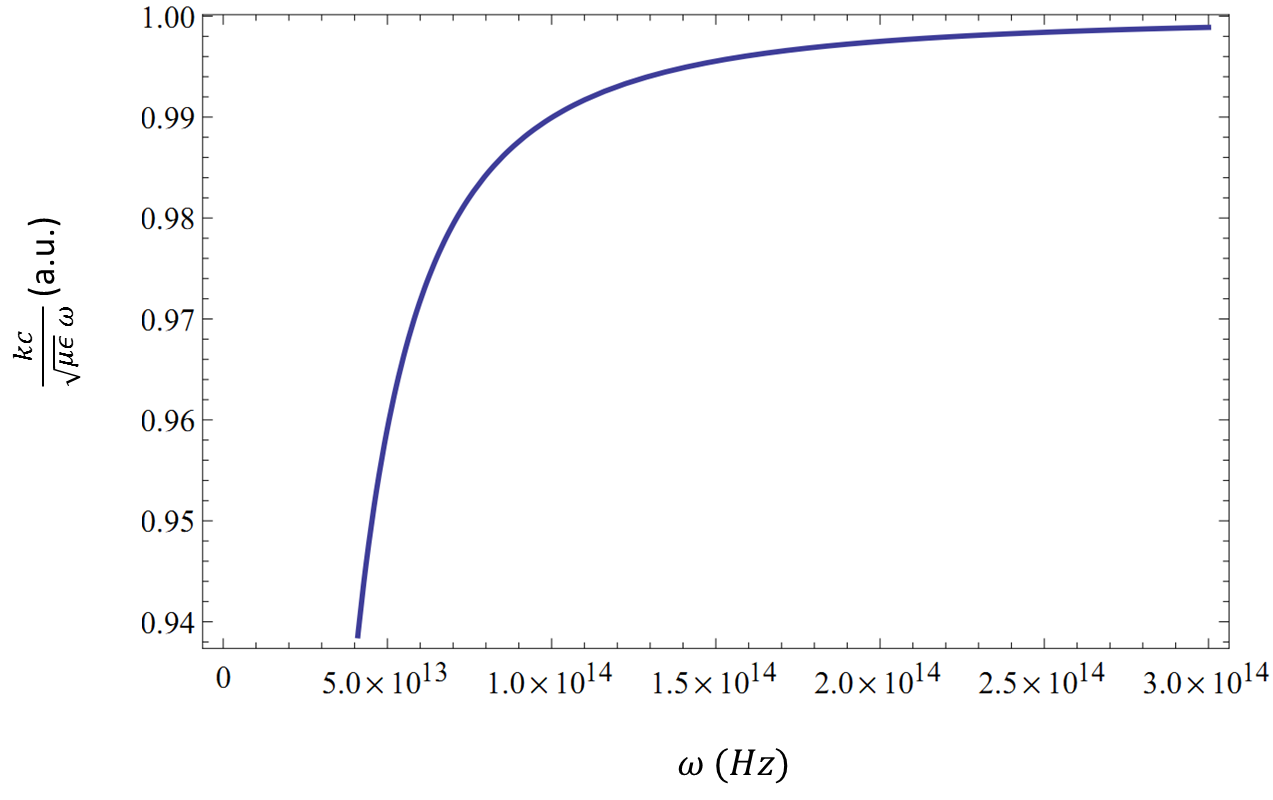}}
\end{center}
\caption{\small Normalized wave vector $\frac{kc}{\sqrt{\mu \epsilon }\omega}$ in arbitrary units (a.u.) as function of $\omega$. We consider $\l=2$ {\rm nm} and $ Bi_2\, Se_3 $ dielectric parameters, $\mu=1$ and $\epsilon=100$ \cite{Hzang}, so the cutoff frequency is $\omega_{\theta} \approx 10^{13}$ {\rm Hz}.} 
\label{fig2}
\end{figure}

Once this slab waveguide confines waves only in $x$ direction, we assume the modes are propagating freely along $z$, so it is necessary to solve the Maxwell equations for planar solutions of EM fields such: $\vec{E}(\vec{r},t)=\vec{E}(x)e^{i(kz-\omega t)}$ and $\vec{B}(\vec{r},t)=\vec{B}(x)e^{i(kz-\omega t)}$. Taking then into Eqs. (\ref{Faraday}) and (\ref{Ampere-Maxwell}) we obtain the transverse fields:
\begin{equation}\label{Ex}
      E_x=  \frac{i}{\mu \epsilon \omega^2-c^2k^2}\left(c^2 k \partial_{x}E_{z}+c \omega \partial_{y}B_{z} +\mu c \omega \theta' (\hat{n} \times \vec{E})_{x} \right),
\end{equation}

\begin{equation}\label{Ey}
      E_y=  \frac{i}{\mu \epsilon \omega^2-c^2k^2}\left(c^2 k \partial_{y}E_{z}-c \omega \partial_{x}B_{z} +\mu c \omega \theta' (\hat{n} \times \vec{E})_{y} \right),
\end{equation}

\begin{equation}\label{Bx}
      B_x=  \frac{i}{\mu \epsilon \omega^2-c^2k^2}\left(c^2 k \partial_{x}B_{z}-\mu \epsilon c \omega \partial_{y}E_{z} +\mu  c^2 k \theta' (\hat{n} \times \vec{E})_{y} \right),
\end{equation}

\begin{equation}\label{By}
      B_y=  \frac{i}{\mu \epsilon \omega^2-c^2k^2}\left(c^2 k \partial_{x}E_{z}+\mu \epsilon c \omega \partial_{y}B_{z} +\mu c^2 k \theta' (\hat{n} \times \vec{E})_{x} \right),
\end{equation}
where $\theta' \equiv \frac{\alpha  \theta}{\pi l}$ and $\hat{n}=\hat{x}$ (see Fig. \ref{fig1}). For the slab waveguide is only necessary to analyze the dynamics over $z$, so, from Eqs. (\ref{Gauss}) and (\ref{no mon}) we obtain:  
\begin{equation}\label{edo Ez}
     \partial_x ^2 E_z(x) + \left(\frac{\mu \epsilon}{c^2}\omega^2 - k^2\right) E_z(x) - {\theta' \over \epsilon}\partial_x B_z(x) -\mu \theta'^2 E_z(x)=0 ,            
 \end{equation} 
     
\begin{equation}\label{edo Bz}
\partial_x ^2 B_z(x) +  \left(\frac{\mu \epsilon}{c^2} \omega^2 -k^2\right) B_z(x) +\mu \theta' \partial_x E_z(x) = 0.                                                                   
\end{equation}
To decouple these equations we must depart to higher orders. Doing that for $B_z(x)$, we get:
\begin{equation}\label{edo Bz4}
\left[\partial_x ^4  + 2\left( {\mu  \epsilon  \over {c^2}}\omega^2- k^2 - {\mu \over \epsilon }\theta'^2\right)\partial_x^2 + \left({\mu  \epsilon   \over {c^2}}\omega^2- k^2 \right)\left({\mu  \epsilon  \over {c^2}}\omega^2- k^2  - {\mu \over \epsilon}{\omega \over{k c}}\theta'^2\right)\right] B_z(x) =0,
\end{equation}\\
where $E_z(x)$ is obtained from Eq. (\ref{edo Bz}). Taking $\theta=0$, we recover the standard equations as expected. The solution has the form: $B_z(x)=B_1 e^{-\gamma_+ x} + B_2 e^{-\gamma_- x}$, valid for the $x>L$, for $x<0$ the exponential signals are inverted so that solution remains finite. The constants $B_1$ and $B_2$ are fixed field amplitudes within the TI, while:
\begin{equation}\label{gamma}
\gamma_{\pm}=\left( k^2 - \frac{\mu \epsilon}{c^2} \omega^2 + {\mu \over \epsilon } \theta'^2  \pm \frac{ \sqrt{ \mu c^5 \theta'^2 k  (2 \epsilon c^3 k^3 + \mu c^3 \theta'^2 k - \epsilon c^2 k^2 \omega - 2 \mu \epsilon^2 c k \omega^2 + \mu \epsilon^2 \omega^3) } }{\epsilon  c^4 k}\right)^{1/2}.
\end{equation}
From the wave equation for EM fields, $\gamma_{\pm}$ must satisfy $\gamma_{-}^2 + \gamma_{+}^2 - k^2 +  \frac{\mu \epsilon}{c^2} \omega^2=0$, so that we finally obtain:
\begin{equation}\label{cutoff}
{k c\over\omega}=\sqrt{\mu\epsilon} \left[1 - \left(\omega_{\theta}\over\omega\right)^{2}\right]^{\frac{1}{2}}, \,\, \omega_{\theta}=\frac{2c}{\pi \epsilon}\frac{\alpha \theta}{l}.
\end{equation}
Waves with $\omega < \omega_{\theta}$ do not propagate in TI surface, since $k$ becomes imaginary resulting in attenuation of the EM radiation. These frequencies propagate along $z$ direction inside the waveguide only, therefore, $\omega_{\theta}$ is the topological cutoff frequency, such behaviour  is illustrated in Fig. \ref{fig2} for $Bi_2\, Se_3$.


Since the cutoff frequency depends on $l$, the system proposed here can be useful to estimate the penetration length of the surface states in 3D TI bulk by measuring $\omega_{\theta}$. The topological nature of $\omega_{\theta}$ stems from the TME on the surface of TI and this frequency corresponds to the minimum energy required for EM radiation through the surface of TI. Thus, EM waves with $\omega < \omega_{\theta}$ are reflected keeping $\rho_{H}$ and $\vec{J}_{H}$ on the slab surface and propagate along the waveguide. These results are in agreement with experimental data reported in Ref.\cite{optical}, which were made measurements of the reflectivity and optical conductivity of four 3D TI (including $ Bi_2\, Se_3 $) from $5$ to $300 K$ and from sub-THz to visible frequencies. The data show a sharp drop in reflectivity ($R(\omega)$) at $\omega \sim 10^{13}$ {\rm Hz} and $R(\omega) \approx 1$ for $\omega < \omega_{\theta}$, as provided in our approach.

\begin{figure}[h]
\begin{center}
\resizebox{!}{6.3cm}{\includegraphics{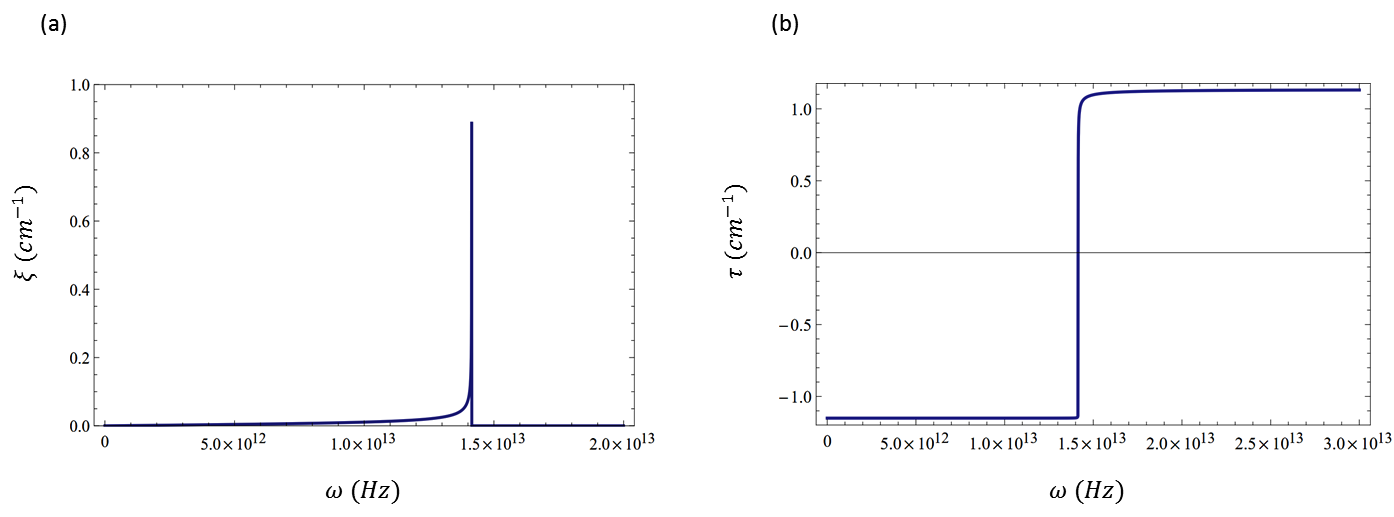}}
\end{center}
\caption{\small Evanescent and oscillatory components of the perpendicular wave vector, $\gamma_{+}(\omega)=\xi (\omega) + i \tau (\omega)$. We take $\l=2$ {\rm nm} then $\omega_{\theta} \approx 10^{13}$ {\rm Hz}. When $\omega=\omega_{\theta}$,  (a) $\xi(\omega)$ vanishes abruptly and (b) for $\tau(\omega)$ there is an abrupt reversal of signal.} 
\label{fig3}
\end{figure}

Along $x$, the wave vector in TI can be expressed as a function of $\omega$ and sweeping the same frequency values made for Eq.  (\ref{cutoff}), we found that Eq. (\ref{gamma}) takes the form: $\gamma_{\pm}(\omega)=\xi(\omega) \pm i\tau(\omega)$. Therefore, the wave vector perpendicular to TI surface have real ($\xi$) and imaginary ($\tau$) values. Typical behaviours of $\xi(\omega)$ and $\tau(\omega)$  are shown in Figs. \ref{fig3} (a) and \ref{fig3} (b) respectively. Below the cutoff frequency, the EM wave is strongly attenuated when it penetrates the TI, once $\xi (\omega)\neq 0$, so when $\omega < \omega_{\theta}$ there is an efficient confinement of the radiation along $x$, according to nontrivial boundary conditions on EM fields (obtained from \ref{Gauss} and \ref{Ampere-Maxwell}):
 \begin{equation}\label{bc}
\left\{ \begin{array}{l}   
\hat{n} \cdot (\vec{E}_{vac}-\epsilon\vec{E}_{TI}) = \rho_{H}\\
\hat{n} \times (\vec{B}_{vac}-\frac{\vec{B}_{TI}}{\mu}) = \vec{J}_{H} ,
\end{array}     \right.
\end{equation}
where subscript $vac$ denote the EM fields in vacuum. In addiction, the stationary wave condition require that inside the wavequide, the wave vector along $x$ ($k_{x}$) is discretized  and depends on the geometrical parameters of the waveguide, i.e., $k_{x}=n \pi/L$ ($n=1,2,3,...$). Due to the large attenuation of EM radiation around $\omega_{\theta}$, the skin depth ($1/\xi$) of EM wave in TI decreases rapidly in $\omega \approx \omega_{\theta}$. For $\omega > \omega_{\theta}$, $\xi(\omega)$ vanishes, while $\tau(\omega)$ reverse signal abruptly resulting in transmission of EM radiation through the walls. This is in accordance whit optical conductivity ($\sigma(\omega)$) measurements performed in Ref.\cite{optical}, where it was observed large values for $\sigma(\omega)$ in this frequency regime.

For the sake of completeness, let us give a brief analysis of transverse polarization modes. The transverse magnetic (TM) mode, $B_{z}(x)=0$, so, from Eqs. (\ref{edo Ez}) and (\ref{edo Bz}) we get: $\partial_x ^2 E_z(x) + \left(\frac{\mu \epsilon}{c^2}\omega^2 - k^2  - \mu \theta'^2 \right)E_z(x)=0$ and $\partial_x E_{z}(x)=0$. Adopting procedure previously applied, we have $\omega_{\theta} \sim 10^{13}$ {\rm Hz} as EM waves with generic polarization ($B_{z}(x) \neq 0$ and $E_{z}(x) \neq 0$). In the transverse electric (TE) polarization, $E_{z}(x)=0$, then: $\partial_x ^2 B_z(x) + \left(\frac{\mu \epsilon}{c^2}\omega^2 - k^2 \right) B_z(x)=0$ and $\partial_x B_{z}(x)=0$. Note that for waveguide operating in the TE mode there is no contribution of topological parameter $\theta$, the transverse components (Eqs. (\ref{Ex})-(\ref{By})) of $\vec{E}$ and $\vec{B}$ fields  vanish on the walls and inside the TI leading to $\rho_{H}=-\frac{\alpha  \theta}{\pi l} (\hat{x} \cdot \vec{B})=-\frac{\alpha  \theta}{\pi l}B_{x}=0$, that is, TE waves do not produce TME in the slab waveguide. With theses conditions, the  cutoff frequencies along the slab are similar to a waveguide with perfectly conducting walls, i.e., $\omega_{m}=m \pi c/L$ $(m=1,2,3,...)$. For EM waves in the TE mode, frequencies less than $\omega_{m}$ do not propagate along the slab waveguide similarly to the conventional waveguides described in Refs. \cite{jack,grif,landau}. 
 
\section{Conclusions} 
 
In summary, we find that waveguide made from 3D TI walls can propagate EM radiation efficiently and have cutoff frequencies with topological stability, which value depends on $l$ of the metallic surface states. The value provided here for $\omega_{\theta}$ agrees with experimental results of optical reflectivity and conductivity of 3D TI reported in Ref.\cite{optical}. This slab waveguide can be useful as a prototype to determine a microscopic quantity, $l$, by measuring a macroscopic one, $\omega_{\theta}$. In the TM mode, the cutoff frequency has the same characteristics and value of the generic polarization. On the other hand, TE mode has no topological properties due to TME not manifest for this polarization state, so the cutoff frequency depends only on the geometric parameter ($L$) similar to waveguide with usual conductive walls. Other geometries as cylindrical waveguide, coaxial cable, resonant cavities, etc. can reveal new and useful electrodynamics properties of 3D TI.

\section*{Acknowledgments}

   The authors thank CNPQ, CAPES and FAPEMIG (brazilian agencies) for financial support.

\end{document}